# Reduction of Threshold Current Density for Current-Driven Domain Wall Motion using Shape Control

Akinobu Yamaguchi, Kuniaki Yano, Hironobu Tanigawa, Shinya Kasai and Teruo Ono

Institute for Chemical Research, Kyoto University, Uji, Kyoto, 611-0011, Japan

We investigated the aspect ratio (thickness/width) dependence of the threshold current density required for current-driven domain wall (DW) motion in $Ni_{81}Fe_{19}$ nanowires. It has been shown theoretically that the threshold current density is proportional to the product of the hard-axis magnetic anisotropy $K_{\perp}$ and the DW width $\lambda$ [G. Tatara and H. Kohno: Phys. Rev. Lett. **92** (2004) 086601.]. We show experimentally that $K_{\perp}$ can be controlled by the magnetic shape anisotropy in the case of $Ni_{81}Fe_{19}$ nanowires, and that the threshold current density increases as $K_{\perp} \cdot \lambda$ increases. We succeeded in reducing the threshold current density by half using the shape control.

# 1. Introduction

Current-driven domain wall (DW) motion has attracted much attention from the viewpoint of applications because this effect makes it possible to switch the magnetic configuration without an external magnetic field[1,2]. The effect has been convincingly confirmed by a series of experiments on magnetic thin films[3-8] and magnetic nanowires[9-20]. However, the threshold current densities required for the current-driven DW motion ($J_C$) are still high, $10^{11} - 10^{12}$ A/m$^2$ for single ferromagnetic metal layer circuits[9-16] and on the order of $10^{10}$ A/m$^2$ for spin-valve ferromagnetic nanowires[17-19]. Although a lower $J_C$ of $10^9$ A/m$^2$ has been reported for the ferromagnetic semiconductor (Ga,Mn)As[20], the Curie temperature of this material is below room temperature. Thus, it is required to explore a way to reduce $J_C$ for ferromagnetic metals from the viewpoint of practical applications.

It has been suggested theoretically that $J_C$ is proportional to the product of the hard-axis magnetic anisotropy $K_\perp$ and the DW width λ in the case of a thick DW and weak pinning[21]. Here, a thick DW means that the thickness of the DW is much larger than the Fermi wavelength of the conduction electrons, which is satisfied in usual ferromagnetic metals, and weak pinning means that the pinning potential for a DW is much smaller than $K_\perp/\alpha$, where $\alpha$ is the Gilbert damping factor. Thus, the theory predicts that $J_C$ can be reduced by reducing $K_\perp$. For samples of Ni$_{81}$Fe$_{19}$, $K_\perp$ can be controlled by the sample shape, because the crystal magnetic anisotropy of Ni$_{81}$Fe$_{19}$ is negligibly small and the magnetic anisotropy is dominated by the magnetostatic energy. In this letter, we show experimentally that $J_C$ depends on the cross-sectional shape of Ni$_{81}$Fe$_{19}$ wires. The effect of the Joule heating of the sample is also discussed.

## 2. Experimental

Samples with two shapes, L-shaped and semicircular-shaped magnetic wires of $Ni_{81}Fe_{19}$, were fabricated on thermally oxidized Si and MgO substrates by electron beam lithography and a liftoff method as shown in Figs. 1(a) and 1(b). We have checked that two types of wires on the same substrate have the same $J_C$ when they have the same thickness and the same width. Samples investigated in this study are summarized in Table 1 with the experimental results. The widths of the wires were determined using a scanning electron microscope, and the thicknesses were determined with an atomic force microscope.

A single DW was introduced into a magnetic wire by the following procedure. For L-shaped wires, the direction of an external magnetic filed was set about $30°$ from the wire axis in the substrate plane in order to introduce DW at a position slightly removed from the corner [Fig. 1(a)]. First, a magnetic field of +2 kOe was applied in order to align the magnetization in one direction along the wire. Then, a single DW was introduced by applying a magnetic field of −175 Oe[13]. In the case of the semicircular-shaped wires, a magnetic field of 20 kOe was applied in the y-direction in the substrate plane, and it was decreased to zero [Fig. 1(b)]. Then, a single DW was introduced spontaneously around the center of the semicircular-shaped wire[22].

The threshold current densities were determined by direct observations of the current-driven DW motion using a magnetic force microscope (MFM) at room temperature[13), 15)]. After the DW introduction, a pulsed current lasting 5 μs long was applied through the wire in the absence of a magnetic field. The density of the pulsed current was increased until the DW was displaced in the direction opposite the pulsed current. CoPtCr low moment probes were used in order to minimize the influence of the stray field from the probe on the DW in the wire.

Magnetic anisotropies of the samples were determined by measuring the magnetoresistance effect. In narrow ferromagnetic wires, the magnetization is restricted to be directed parallel to the wire axis due to the magnetic shape anisotropy. When a magnetic field is applied perpendicular to the wire axis, the magnetization is tilted from the wire axis, and the angle between the magnetization and the measuring electric current increases. This causes a decrease in resistance, and the resistance is minimized value when the magnetization is directed parallel to the external magnetic field, because $Ni_{81}Fe_{19}$ shows the anisotropic magnetoresistance (AMR) effect. Thus, magnetic field at saturation can be determined by the magnetoresistance measurement. We refer to the saturation fields along the two magnetic hard axes as $H_{S//}$ and $H_{S\perp}$, respectively, as shown schematically in Fig. 1(c).

## 3. Results and Discussion

The results of the magnetoresistance measurements at 300 K for samples #1, #2, #3, and #4 in Table 1 are shown in the inset of Fig. 2. The magnetic field was applied perpendicular to the substrate plane. As indicated by the arrows, $H_{S\perp}$ decreases monotonically as the thickness of the wire increases. It should be noted that the sum of $\mu_0 H_{S//}$ and $\mu_0 H_{S\perp}$ for each sample was about 1.1 T, which is the saturated magnetization of $Ni_{81}Fe_{19}$ ($M_S$). This suggests that the shape anisotropy dominates the magnetic anisotropy in the samples. Because $H_{S//}$ in the semicircular-shaped wire can not be determined by measuring the magnetoresistance effect under a magnetic field in the plane because of its shape, we defined $\mu_0 H_{S//}$ as the difference between the saturated magnetization and $\mu_0 H_{S\perp}$, $\mu_0 H_{S//} = M_S - \mu_0 H_{S\perp}$. This can be justified because the radius of the semicircular-shaped wire is much larger than both the thickness and the width of the wires. The hard-axis magnetic anisotropy $S^2 K_\perp / a^3$ in ref. 21 is expressed by experimentally obtained values as

$$\frac{S^2 K_\perp}{a^3} = M_S \cdot \left| H_{S\perp} - H_{S//} \right| \quad (1)$$

where $S$ and $a$ are the localized spin and the lattice constant, respectively. Figure 2 shows $S^2 K_\perp / a^3$ as a function of the aspect ratio (thickness/width). $S^2 K_\perp / a^3$ decreased systematically as the aspect ratio increased. MFM observations under a magnetic field revealed that the samples had depinning fields in the range from 15 to 100 Oe, suggesting that the pinning potentials in the wires were much smaller than $K_\perp / \alpha$ and that the samples were in the weak pinning regime.

The high current density required for the current-driven domain wall motion inevitably causes considerable Joule heating as previously reported[23]. We estimated the sample temperature during the application of the pulsed-current using the method described in ref. 23. Figure 3 shows the estimated temperatures for samples #6 and #9 as a function of current density. The shape of these two wires was almost the same, but the estimated temperature of the wire on the thermally oxidized Si substrate (sample #6) was higher than that of the wire on the MgO substrate (sample #9) at the same current density, because MgO has a higher thermal conductivity than $SiO_2$. $J_C$ for each sample is indicated by the arrows in Fig. 3. $J_C$ of the wire on the thermally oxidized Si substrate was much smaller than that of the wire on the MgO substrate in spite of the similar dimensions of the two wires. This can be attributed to the reduction in $M_S$ due to the higher sample temperature at $J_C$ for the sample on the thermally oxidized substrate. The reduction of $M_S$ results in the decrease of $S^2 K_\perp / a^3$ through eq. (1), leading to the reduction of $J_C$. This indicates that we should take into account the decrease in the magnetization at saturation due to Joule heating.

Another indication of the importance of sample heating is seen in Fig. 4(a) where $J_C$ is plotted as a function of $S^2 K_\perp \cdot \lambda / a^3$. Here, the DW width, $\lambda$, was calculated using the micoromagnetic simulation The Objected Oriented MicroMagnetic Framework (OOMMF)[24]. As shown in Fig. 4(a), $J_C$ values for samples #5 and #6 are smaller than that for the sample #9, although the values of $S^2 K_\perp \cdot \lambda / a^3$ of #5 and #6 are larger than that of #9. This result is contrary to the theoretical suggestion that larger $S^2 K_\perp \cdot \lambda / a^3$ leads to higher $J_C$. This discrepancy can be resolved by taking into account that the $K_\perp$ is reduced because of the reduction in $M_S$ due to sample heating as described below. Figure 4(b) shows $J_C$ as a function of $S^2 K_{\perp eff} \cdot \lambda / a^3$, which was calculated using eq. (1) with the reduced magnetization at $J_C$. In this plot, $J_C$ shows systematic behavior vs. $S^2 K_{\perp eff} \cdot \lambda / a^3$. $J_C$ increases with $S^2 K_{\perp eff} \cdot \lambda / a^3$ and is minimized for $S^2 K_{\perp eff} \cdot \lambda / a^3 = 0$, which is about half of the previously reported value for 10-nm-thick $Ni_{81}Fe_{19}$ wire 240 nm wide[13), 23)]. The tendency that $J_C$ increases with $S^2 K_{\perp eff} \cdot \lambda / a^3$ is qualitatively consistent with theory[21], indicating that spin-transfer is a most probable mechanism. However, theory predicts much larger values of $J_C$ than the experimental ones. One possible reason for this discrepancy between theory and experiment is the internal spin structure of a DW; the theory assumes a simple one-dimensional DW, while the DW in the experiments has a complicated internal spin structure as previously reported[10), 13)]. Another possibility is the existence of the field-like term as suggested by recent theories[25-28)].

## 4. Conclusions

We have shown that the threshold current density can be decreased by reducing the effective $K_\perp$. The smallest threshold current density was about half of the previously reported value for 10-nm-thick $Ni_{81}Fe_{19}$ wire 240 nm wide[13), 23)], and this shows the effectiveness of shape control. It was also suggested that the decrease in the magnetization at saturation by sample heating plays a role in the reduction of the threshold current density.

## Acknowledgements

The authors are grateful to G. Tatara, H. Kohno, and Y. Nakatani for valuable discussions. This work was partly supported by Ministry of Education, Culture, Sports, Science and Technology Grants-in-Aid for Scientific Research in Priority Areas, JSPS Grants-in-Aid for Scientific Research, and Industrial Technology Research Grant Program in '05 from NEDO of Japan.

**Figure caption**

Figure 1(a) Schematic illustration of the top view of a L-shaped wire. (b) Schematic illustration of the top view of a semicircular-shaped wire. (c) The magnetic easy axis is parallel to the wire axis. Two magnetic hard axes are perpendicular to the wire axis. One is in the substrate plane, and the other is perpendicular to the substrate plane.

Figure 2 The hard-axis magnetic anisotropy $S^2 K_\perp / a^3 = M_s \cdot |H_{s\perp} - H_{s//}|$ is plotted as a function of the aspect ratio (thickness/width). The solid squares, the solid circles, and the solid triangles indicate the results for the semicircular-shaped wires on MgO substrate, the L-shaped wires on the thermally oxidized Si substrates, and the semicircular-shaped wires on the thermally oxidized Si substrates, respectively. The inset shows the results of magnetoresistance measurements at 300 K for the samples #1, #2, #3, and #4. The magnetic field was applied perpendicular to the substrate plane. The magnetization fields at saturation are indicated by the arrows.

Figure 3 The estimated temperatures for samples #6 and #9 are plotted as a function of current density.

Figure 4 (a) The experimentally determined $J_C$ is plotted as a function of $S^2 K_\perp \cdot \lambda / a^3$. (b) The experimentally determined $J_C$ is plotted as a function of $S^2 K_{\perp eff} \cdot \lambda / a^3$, which was calculated using eq. (1) with the reduced magnetization at $J_C$. The solid squares, the solid circles, and the solid triangles indicate the results for the semicircular-shaped wires on the MgO substrate, the L-shaped wires on the thermally oxidized Si substrates, and the semicircular-shaped wires on the thermally oxidized Si substrates, respectively.

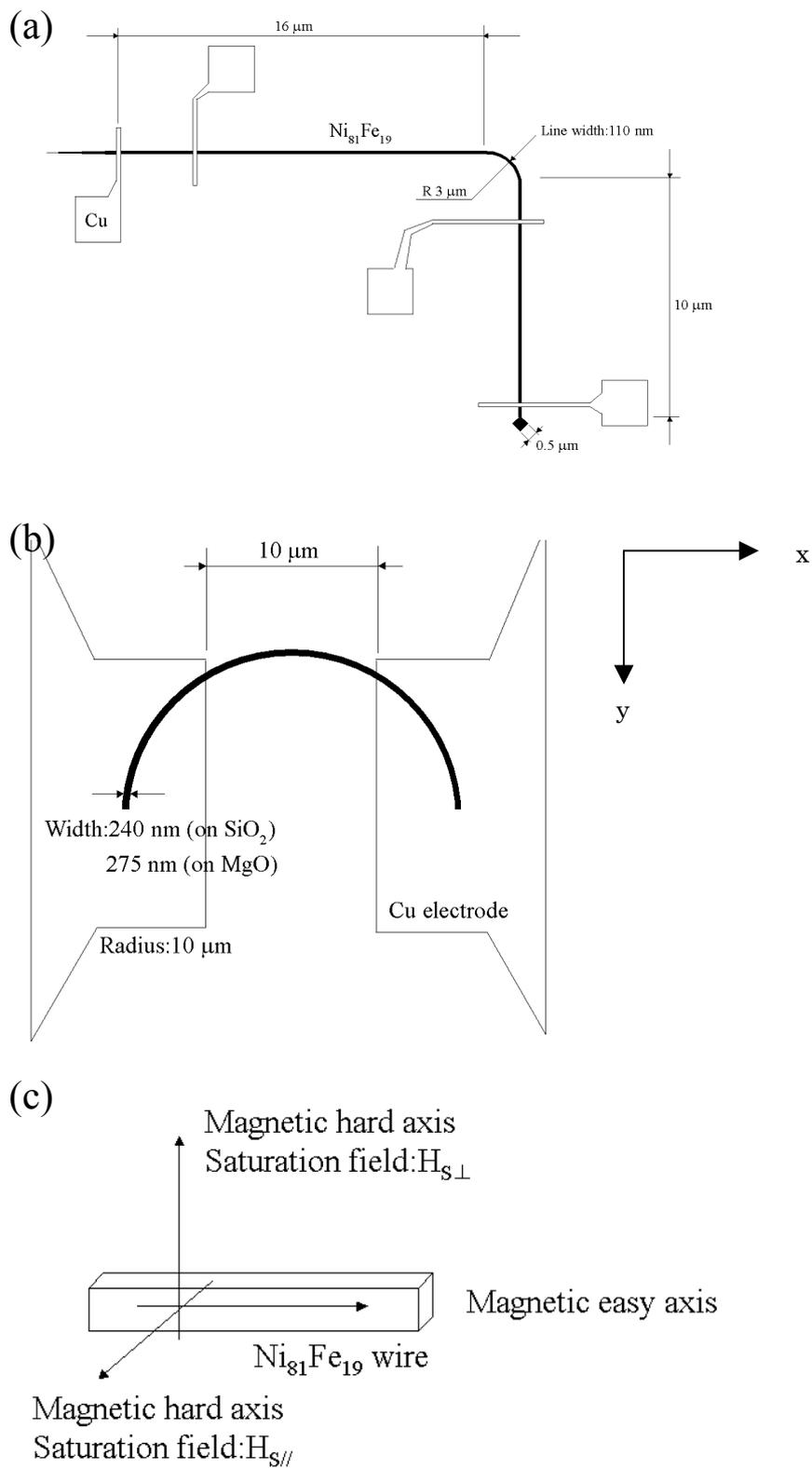

Fig. 1_Yamaguchi_et_al.

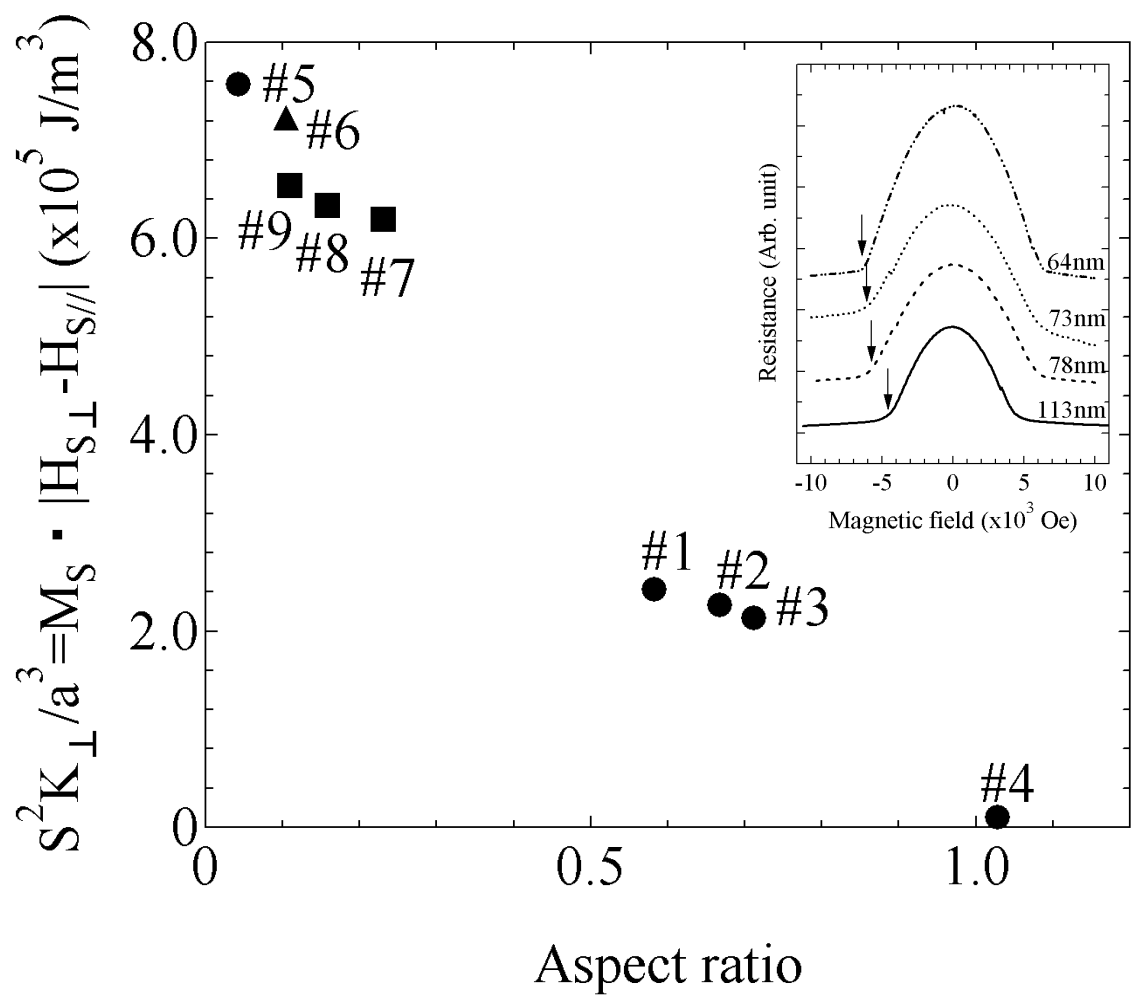

Fig. 2_Yamaguchi_et_al.

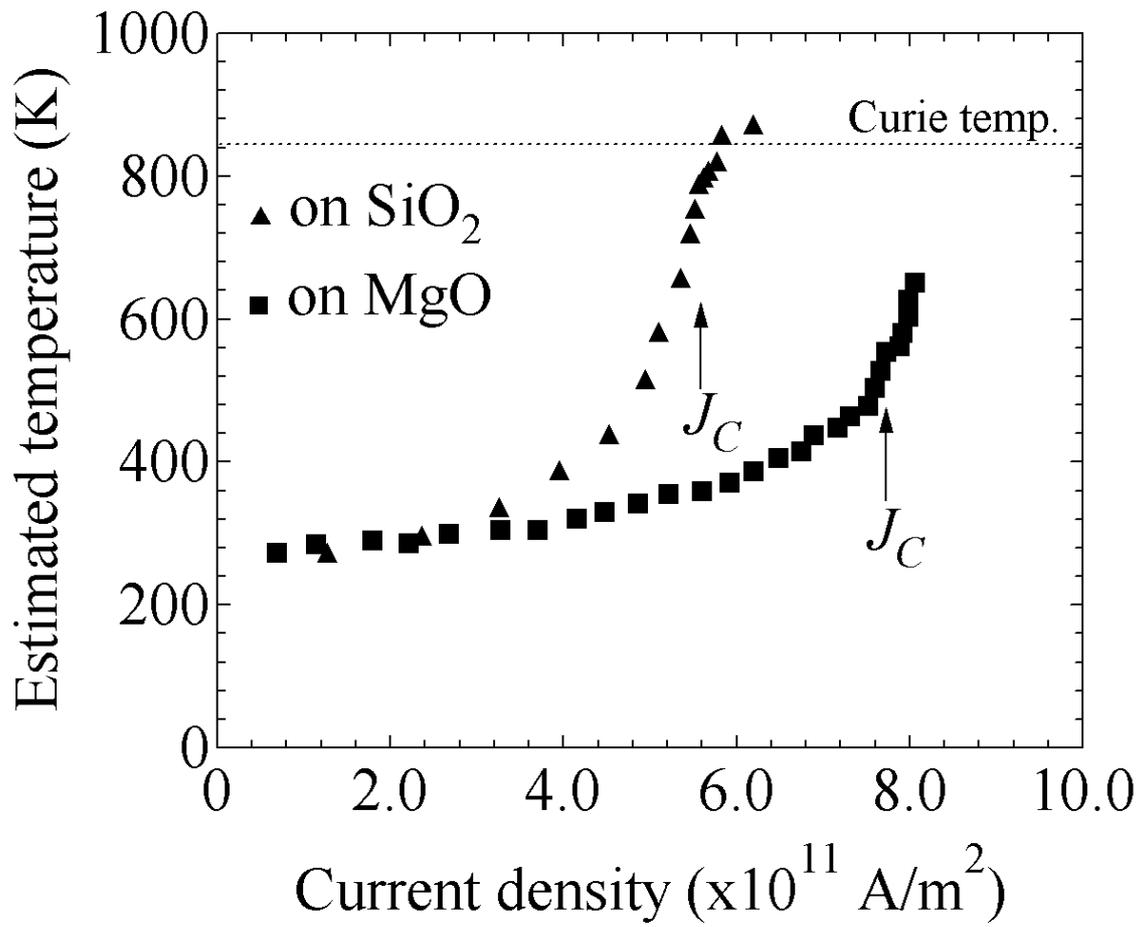

Fig. 3_Yamaguchi_et_al.

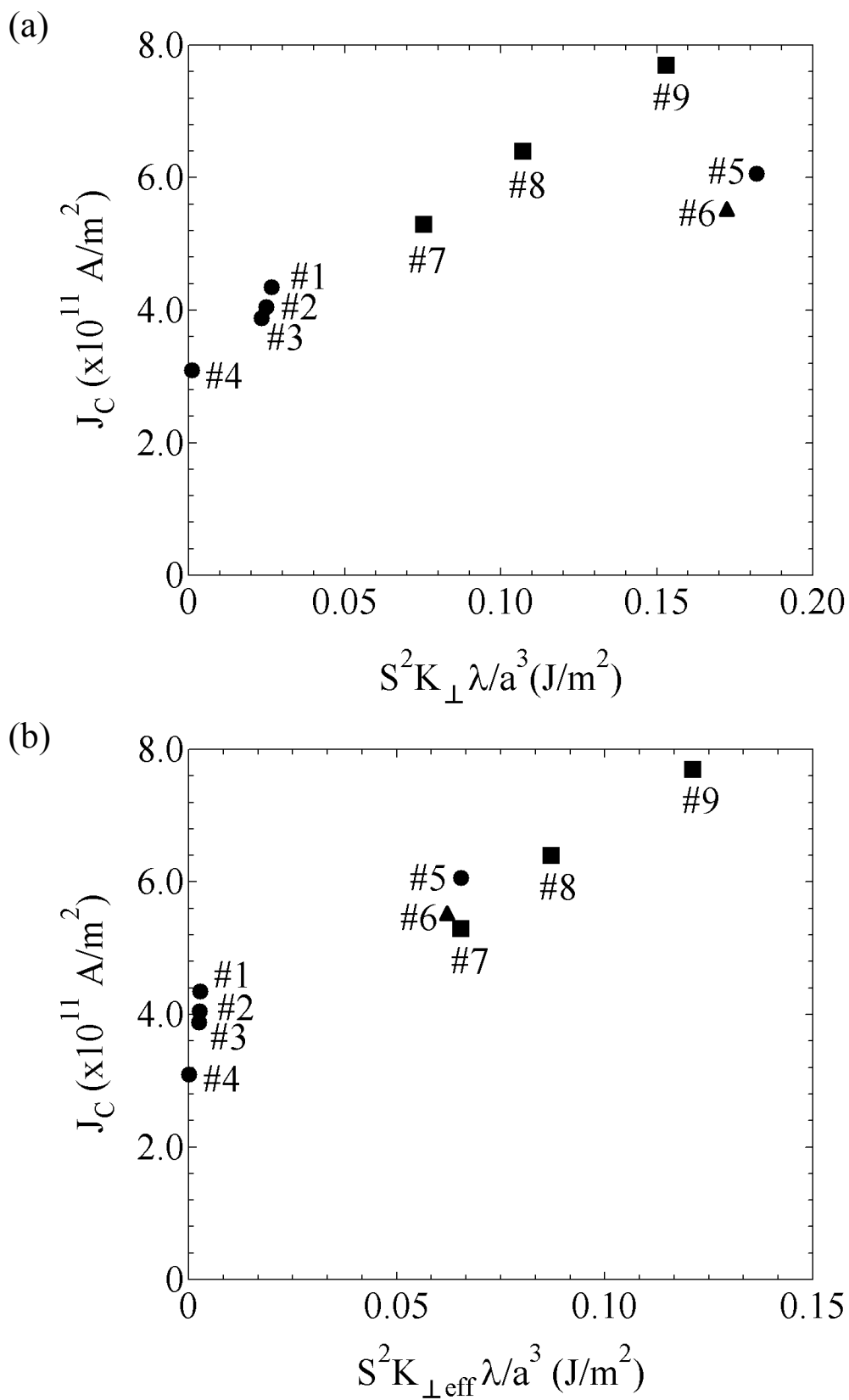

Fig. 4_Yamaguchi_et_al.

Table I. Summary of the samples and the experimental results.

| Sample | Substrate | Type* | Width (nm) | Thickness (nm) | Temperature** (K) | $M/M_s$** | $J_C$ ($\times 10^{11}$ A/m$^2$) |
|---|---|---|---|---|---|---|---|
| #1 | SiO$_2$ | L | 110 | 64 | 820 | 0.33 | 4.34 |
| #2 | SiO$_2$ | L | 110 | 73 | 820 | 0.33 | 4.04 |
| #3 | SiO$_2$ | L | 110 | 78 | 820 | 0.33 | 3.88 |
| #4 | SiO$_2$ | L | 110 | 113 | 820 | 0.33 | 3.09 |
| #5 | SiO$_2$ | L | 240 | 10 | 750 | 0.60 | 6.06 |
| #6 | SiO$_2$ | C | 240 | 25 | 750 | 0.60 | 5.50 |
| #7 | MgO | C | 130 | 30 | 400 | 0.90 | 5.30 |
| #8 | MgO | C | 190 | 30 | 520 | 0.85 | 6.40 |
| #9 | MgO | C | 275 | 30 | 550 | 0.82 | 7.70 |

*Type L and C correspond to the L-shaped and semicircular-shaped wires, respectively.

**Temperature and magnetization at the threshold current density.